\documentclass[aps,pra,twocolumn]{revtex4}
\usepackage{amsfonts}
\usepackage{bbding}
\usepackage{amssymb}
\usepackage{color}
\usepackage{amsmath}
\usepackage{graphicx}
\usepackage{subfigure}
\usepackage{txfonts}
\usepackage{bm}
\usepackage{ulem}
\usepackage[colorlinks,linkcolor=blue,citecolor=blue]{hyperref}
\newcommand{\jj}[1]{{\color{black}#1}}
\newcommand{\xl}[1]{{\color{black}#1}}

\begin{document}

\title{Steady-state susceptibility in continuous phase transitions of dissipative systems}
\author{Xingli Li, Yan Li, and Jiasen Jin}
\email{jsjin@dlut.edu.cn}
\affiliation{School of Physics, Dalian University of Technology, 116024 Dalian, China}
\date{\today}

\begin{abstract}
In this work, we explore the critical behaviors of fidelity susceptibility and trace distance susceptibility associated to the steady states of dissipative systems at continuous phase transitions. We investigate on two typical models, one is the dissipative spin-1/2 XYZ model on two-dimensional square lattice and the other is a driven-dissipative Kerr oscillator. We find that the susceptibilities of fidelity and trace distance exhabit singular behaviors near the critical points of phase transitions in both models. The critical points, in thermodynamic limit, extracted from the scalings of the critical controlling parameters to the system size or nonlinearity agree well with the existed results.
\end{abstract}

\maketitle
\section{Introduction}
\label{Introduction}
Phase transition is the key concept in condensed matter and statistical physics \cite{SSachdev}. In general, the quantum phase transitions in equilibrium case always signify that the ground-state properties of the quantum many-body systems have changed, in particular, the correlation length, magnetic susceptibility, entanglement and other physical quantities exhibit the divergent behaviors at the critical point \cite{Amico2008RMP}. However, a realistic system is always regarded as open system due to the inevitable interactions with its environment  \cite{Lindblad1976,Gorini1976,breuer_book}. The dissipation induced by the system-environment interactions drive the open system away from the equilibrium. The phase transition may also occur in open quantum many-body system manifested by the emerging of ordered steady state in the long-time limit when tuning the controllable parameter \cite{Henkel_book1,Henkel_book2}. Investigating such nonequilibrium phase transitions is not only of great significance in the understanding the collective phenomena and dynamics of dissipative systems, but also highlighting the possibilities of quantum information processing in open systems, such as quantum state engineering and quantum sensing.

Due to the nonunitary nature of the dynamics and exponential growth of the dimension of Hilbert space of the system, it is challenge in the theoretic description and numerical simulation for the open quantum many-body systems. In recent years, a plenty of meaningful results have been obtained \cite{JiasenJin2016PRX,Biella2018PRB,RiccardoRota2019PRL,Maghrebi2016PRB,overbeck2017,Huybrechts2020PRB,kilda2021} and several numerical methods have been developed \cite{TonyELee2013PRL,TonyELee2011PRA,Rigol2006PRL,Tang2013CPC,SFinazzi2015PRL,weimer2015,kshetrimayum2017,mckeever2021,huber2021,singh2021} in recent years. By combining the corner-space renormalization method \cite{SFinazzi2015PRL} with the Monte Carlo wave-function trajectory method \cite{Dalibard1992PRL}, Rota {\it et. al.} investigated the divergence of the angularly averaged magnetic susceptibility in the two-dimensional dissipative spin-1/2 XYZ model on square lattice.
With the finite-size scaling analysis, they obtained the critical point of the phase transition from paramagnetic to ferromagnetic phases. The corresponding critical exponents are obtained as well. Besides the angularly averaged magnetic susceptibility and von Neumann entropy, they found that the multipartite entanglement witnessed by the quantum Fisher information also exhibits the critical behavior when the controlling parameter is close to the critical point \cite{Rota2017PRB}. An alternative way to uncover the dissipative phase transition is to investigate the system through the dynamical behavior which is related to the Liouvillian spectrum. The Liouvillian gap tends to close when the system approaches to the critical point, as a consequence the system takes longer time to reach the steady state. The critical slowing down has been observed in two-dimensional driven-dissipative Bose-Hubbard model  \cite{Vicentini2018PRA} and dissipative spin-1/2 XYZ model \cite{Riccardo2018NJP}. Meanwhile, for the latter model, the Liouvillian gap saturates for one-dimensional case signaling the absence of the dissipative phase transition.

Different from the previous studies on the dissipative quantum many-body system, in this work we mainly focus on an observable-independent way to estimate the critical point, which is derived from the response degree of the steady states of the system to the parameter perturbations. To be more precise, we determine the critical point by intuitively giving the degree of similarity of the steady states of the system before and after the perturbation. The basic idea is that, in the thermodynamic limit, when the dissipative systems undergo the phase transitions by tuning the controlling parameter of the Hamiltonian, not only the steady-state order parameters belong to different phases change abruptly, but also the similarity between the two steady states shows a significant dip in the vicinity of the phase transition, such as the fidelity may exhibits divergent behavior in phase transition of the closed system \cite{GuIJMPB2010,ZanardiPRE2006}. However, because the dimension of the Hilbert space in a practical simulations is limited by the computational power, the singular behavior of the similarity of steady states in different phases is usually not revealed. Here we utilize the fidelity susceptibility $\chi_{F}$ as an indicator to uncover the singular behavior.

The fidelity susceptibility, which is a higher-order derivative of fidelity, originates from the linear response theory and differential-geometric approach \cite{TonchevPRE2014}. It is sensitive in reflecting the stability of a given system to the parameter perturbation. Actually, the fidelity susceptibility has already been widely used in characterizing the ground-state phase transition in equilibrium. The non-analytic divergence behavior of the fidelity susceptibility have been observed in the phase transitions of first and second orders \cite{GushijianPRE2008,GushijianPRB2008,AlbuquerquePRB2010,rossini2018,GushijianPRE2007}. Moreover, the fidelity susceptibility have also been used to investigate the nonzero temperature phase transitions \cite{wangleiPRX2015} and topological phase transitions \cite{YangPRA2008,zhaoArxiv2008}.

In this paper, inspired by the good performance of fidelity susceptibility in characterizing the equilibrium phase transitions, we employ the fidelity susceptibility $\chi_{F}$ and trace distance susceptibility $\chi_{T}$ to investigate the dissipative phases transitions in open quantum many-body systems. We determine the critical points by the finite-size scaling analysis on both $\chi_{F}$ and  $\chi_{T}$ for the dissipative spin-1/2 XYZ model on two-dimensional square lattice in which the existence of a steady-state phase transition from the normal paramagnetic to the ordered ferromagnetic phases has been verified by many numeric methods. Our results show that both quantities, which are observable independent, exhibit non-analytic singular behaviors at the vicinity of the phase transitions. We also investigate the system of a driven-dissipative Kerr oscillator.  We verify the existence of the continuous steady-state phase transition through the semiclassical approximation and obtain the similar results near the phase transition.

This paper is organized as follows: In Sec.\ref{Theoretical framework}, we introduce the theoretical framework of the fidelity susceptibility and the trace distance susceptibility for mixed state. In Sec.\ref{ModelI}, we investigate the the critical behavior of the two-dimensional spin-1/2 XYZ model on a square lattice and determine the critical exponents. In Sec.\ref{ModelII}, we employ the semiclassical approximation as a preliminary exploration and then identify the occurrence of dissipative phase transitions in driven-dissipative Kerr oscillator model. We summarize in Sec.\ref{Summary}

\section{THEORETICAL FRAMEWORK}
\label{Theoretical framework}
In this section, we introduce the theoretic descriptions for the dynamics of open quantum many-body systems and the concepts of fidelity susceptibility and trace distance susceptibility. We focus on the quantum many-body systems that are subjected to local environments. Under the Markovian approximation, the dynamics of the system's density matrix can be described by the Lindblad master equation ($\hbar=1$ hereinafter)
\begin{equation}
\frac{\partial\hat{\rho}}{\partial t} = \mathcal{L}\hat{\rho}(t) = -i[\hat{H},\hat{\rho}] + \sum_{j}\mathcal{D}_{j}[\hat{\rho}],
\label{MasterEquation}
\end{equation}
where $\mathcal{L}$ is the non-Hermitian Liouvillian superoperator, $\hat{H}$ is the Hamiltonian of the many-body system, and the dissipator $\mathcal{D}_{j}[\hat{\rho}]$ rules the interplay between the system and the local external environments. The first term on the right-hand-side of Eq. (\ref{MasterEquation}) describes the coherent time-evolution that ruled by the Hamiltonian, while the second term describes the incoherent dissipation due to the system-environment interactions.

In general, the eigenspectrum of superoperator $\mathcal{L}$ is complex. The eigenvalue equation is given by,
\begin{equation}
\mathcal{L}\hat{\rho}_{i} = \lambda_{i}\hat{\rho}_{i},
\label{EigEqL}
\end{equation}
where $\lambda_i$ ($i=0,1,2,...$) and $\hat{\rho}_i$ are the eigenvalues and (normalized) eigenstates of $\mathcal{L}$, respectively. Usually the eigenvalues are sorted by the real parts as $\text{Re}[\lambda_{0}]>\text{Re}[\lambda_{1}]>\text{Re}[\lambda_{2}]>\cdots$. The real parts of the eigenvalues are \xl{negative semi-definite}. There is always at least one zero eigenvalue and the associated eigenstate is considered to be the steady state which is denoted by $\hat{\rho}_{\text{ss}}=\hat{\rho}_0$  (the subscript ``ss" denotes steady state). This can be understood as the following. Suppose that the system is initialized in the state $\hat{\rho}(0)=\sum_i{c_i\hat{\rho}_i}$ where $c_i$ are the probability amplitudes. According to Eq. (\ref{EigEqL}), the state of system at arbitrary time evolves to $\hat{\rho}(t)=\sum_i{e^{\lambda_it}c_i\rho_i}$. Apparently, after sufficient long time, all the eigenstates disappear asymptotically except for $\hat{\rho}_{0}$. Moreover, the eigenvalue with the largest nonzero real part is defined as the Liouvillian gap or the asymptotic decay rate \cite{Kessler2012}. The associated eigenstate decays slowest. By tuning the controllable parameter $p$ in the Hamiltonian, the Liouvillian gap may start to close at a critical point $p_c$ indicating the occurrence of the continuous steady-state phase transition in open quantum many-body systems \cite{Minganti2018}.

The steady-state phases are characterized by the order parameter $\langle\hat{O}\rangle_{ss} = \text{Tr}(\hat{\rho}_{ss}\hat{O})$ which is the expected value of an appropriate observable $\hat{O}$ in the steady state. The nonzero order parameter indicates the ordered steady-state phase. In particular, the $M$-order phase transition can be defined as \cite{Minganti2018},
\begin{equation}
\lim_{p\to p_{c}}\Big|\frac{\partial^{M}}{\partial p^{M}} \langle\hat{O}\rangle_{ss}\Big| \rightarrow +\infty.
\label{M-orderPT}
\end{equation}
The discontinuity of the $M$-order derivatives of the order parameter implies that the properties of the steady states are dramatically changed.
It should be noted that the observable $\hat{O}$ is parameter $p$-independent which means that the singularity in Eq.(\ref{M-orderPT}) stems from the steady-state density matrix itself. This reminds us that the abrupt change of the similarity between the states associated to two close parameters may signal the occurrence of phase transition.

In thermodynamic limit, if the Hamiltonian at the critical point is perturbed to $\hat{H}(p_c)\rightarrow\hat{H}(p_c+\delta p)$, the perturbed Hamiltonian can be expressed as
\begin{equation}
\hat{H}(p_{c}+\delta p)=\hat{H}(p_{c})+\hat{H}(\delta p).
\end{equation}
The corresponding steady states are denoted by $\hat{\rho}_{ss}(p_{c})$ and $\hat{\rho}_{ss}(p_{c}+\delta p)$, respectively. When the parameter perturbation drives the system go across the critical point, the properties of the steady states belong different phases will change remarkably. Along this line, quantifying the differences between two steady states in different phases may help us to determine the critical point.

The \textit{fidelity} quantifies the overlap between two given quantum states. It was introduced to characterize the response of a quantum system to a perturbation \cite{PeresPRA1984}. The fidelity is a non-negative, continuous, and symmetric function and it is invariant under unitary transformation \cite{SommersPRA2005}. The fidelity between the steady states associated to $\hat{H}(p)$ and $\hat{H}(p+\delta p)$ in Hamiltonian is given by Uhlmann as the following \cite{Uhlmann1976,TonchevPRE2014},
\begin{equation}
F(p,p+\delta p) = \text{Tr}\sqrt{\sqrt{\hat{\rho}_{ss}(p)}\hat{\rho}_{ss}(p+\delta p)\sqrt{\hat{\rho}_{ss}(p)}}.
\label{fidelity}
\end{equation}

For sufficiently small perturbation, one can expand Eq. (\ref{fidelity}) in powers of $\delta p$,
\xl{
\begin{equation}
F(p,p+\delta p)_{\delta p\to0} \simeq 1 - \frac{\chi_F(p,p+\delta p)}{2}\delta p^2 ...+ O(\delta p^{n}).
\end{equation}}
The \textit{fidelity susceptibility} is defined by the coefficient of the quadratic term $\chi_{F}$ which characterizes the response of fidelity to the parameter $p$. The definition of $\chi_{F}$ is rooted in the Bures distance between two infinitesimally close density matrices \cite{SommersJPA2003}, the expression is governed by
\cite{DSafranekPRA2017}

\begin{equation}
\chi_F(p,p+\mathrm{d}p) = \frac{1}{2}\sum_{n, m}\frac{|\langle m|\delta \hat{\rho}| n\rangle|^{2}}{\lambda_{m}+\lambda_{n}},
\end{equation}
where $\delta \hat{\rho} = \hat{\rho}_{ss}(p + \delta p) -\hat{\rho}_{ss}(p)$ and $|n\rangle$ is the eigenstate associated to the eigenvalue $\lambda_{n}$ of the steady state $\hat{\rho}_{ss}(p)$. In brief, the fidelity susceptibility is the higher-order derivative of the fidelity, which means this quantity is much more sensitive to perturbation.

There are also alternative quantities can be used to measure the difference between two density matrices. For example, the trace distance measures the distance of two quantum states in Hilbert space which satisfies the non-negative definiteness, homogeneity, and the triangle inequality \cite{Wildeurl}. Hence, different from the quantum fidelity, the trace distance is a real distance in Hilbert space. For two arbitrary steady states $\hat{\rho}_{ss}(p_{A})$ and $\hat{\rho}_{ss}(p_{B})$, the trace distance is originally defined as follows,
\begin{equation}
\begin{aligned}
T(p_{A},p_{B}):=&\frac{1}{2}||\hat{\rho}_{ss}(p_{B})-\hat{\rho}_{ss}(p_{A})||_{1} \\
= &\frac{1}{2}\text{Tr}[\sqrt{(\hat{\rho}_{ss}(p_{B})-\hat{\rho}_{ss}(p_{A}))^\dagger(\hat{\rho}_{ss}(p_{B})-\hat{\rho}_{ss}(p_{A}))}].
\end{aligned}
\end{equation}
Consider two steady states as $ \hat{\rho}_{ss}(p_{A}) = \hat{\rho}_{ss}(p)$ and $\hat{\rho}_{ss}(p_{B})= \hat{\rho}_{ss}(p+\delta p)$ with $\delta p$ being the parameter perturbation, we obtain a more refined form of the trace distance as follows,
\begin{equation}
T(p_{A},p_{B})= \frac{1}{2}\text{Tr}[\sqrt{\delta\hat{\rho}^\dagger\delta\hat{\rho}}],
\end{equation}
where $\delta p = \hat{\rho}_{ss}(p + \delta p) -\hat{\rho}_{ss}(p)$. Now we can define the \textit{trace distance susceptibility} as follows,
\begin{equation}
\chi_T(p + \delta p,p) = \frac{1}{2}\text{Tr}\left[\sqrt{\delta\hat{\rho}^\dagger\delta\hat{\rho}}\right]/\delta p.
\end{equation}

\section{Results}
In this section, we investigate both the fidelity susceptibility and trace distance susceptibility in two specific models. We analyze the their scaling behaviors when the controllable parameters go across the steady-state phase transitions to extract the information of the critical points.

\subsection{The dissipative spin-$1/2$ XYZ model}
\label{ModelI}
We start with the model of spin-1/2 particles on the square lattice in two dimension. The Hamiltonian of the many-body system is given by the anisotropic Heisenberg interactions as the following,
\begin{equation}
\hat{H} = \sum_{\langle j,l\rangle}J_{x}\hat{\sigma}^{x}_{j}\hat{\sigma}^{x}_{l} + J_{y}\hat{\sigma}^{y}_{j}\hat{\sigma}^{y}_{l} + J_{z}\hat{\sigma}^{z}_{j}\hat{\sigma}^{z}_{l},
\label{Hamiltonian}
\end{equation}
where $\hat{\sigma}^{\alpha}_{j}(\alpha = x,y,z)$ are the Pauli matrices for the $j$th site, $\langle j,l\rangle$ denotes the nearest-neighbors (NN) interactions and the $J_\alpha$ are the coupling strength. In addition, we assume that each spin couples with a Markovian bath individually which tends to incoherently flip the spin down to the $z$ direction. Under the Born-Markovian approximation and the secular approximation, the dissipator in Eq.(\ref{MasterEquation}) is given by
\begin{equation}
\sum_{j}\mathcal{D}_{j}[\hat{\rho}] = \frac{\gamma}{2}\sum_{j}[2\hat{\sigma}^{-}_{j}\hat{\rho}\hat{\sigma}^{+}_{j} - \{\hat{\sigma}^{+}_{j}\hat{\sigma}^{-}_{j},\hat{\rho}\}],
\label{Lindbladian}
\end{equation}
where $\hat{\sigma}^{\pm}_{j} = (\hat{\sigma}^{x}_{j}\pm i\hat{\sigma}^{y}_{j})/2$ are the raising and lowering operator, respectively, $\gamma$ is the decay rate and $\{\cdot,\cdot\}$ denotes the anticommutator. The master equation admits a $\mathbb{Z}_{2}$ symmetry which means that the system is invariance after a $\pi$-rotation of all spins along the $z$-axis $(\hat{\sigma}^{x}_{j}\to-\hat{\sigma}^{x}_{j},\hat{\sigma}^{y}_{j}\rightarrow-\hat{\sigma}^{y}_{j}, \forall j)$. \xl{Hereinafter we set $\gamma=1$.}

Generally, in the thermodynamic limit, all the spins in steady states will point down along the $z$-axis with zero magnetization the $xy$ plane which is named as paramagnetic phase. However, as the coupling strength varying, the steady states of the system may undergo a phase transition to the ordered phase with nonzero magnetization on the $xy$ plane which is referred to as the ferromagnetic phase implying the spontaneous broken the $\mathbb{Z}_{2}$ symmetry. The steady-state phase transitions in dissipative spin-1/2 XYZ model has been widely studied by many literatures \cite{TonyELee2013PRL,JiasenJin2016PRX,Biella2018PRB,Riccardo2018NJP,Rota2017PRB,kilda2021}.

Here we briefly review the steady-state properties of the considered model. By means of the Gutzwiller mean-field factorization, the density matrix for the total system can be factorized as the tensor product of the density matrices of each site, $\hat{\rho} = \bigotimes_j \hat{\rho}_j$. Each site are assumed to be identical. As a consequence, Eq. (\ref{MasterEquation}) is then reduced to the single-site master equation for a single-qubit system as the following,
\begin{equation}
\frac{d\hat{\rho}}{dt} = -i[\hat{H}^{\text{mf}},\hat{\rho}]+\frac{\gamma}{2}[2\hat{\sigma}^{-}\hat{\rho}\hat{\sigma}^{+} - \{\hat{\sigma}^{+}\hat{\sigma}^{-},\hat{\rho}\}],
\label{MFmasterequation}
\end{equation}
where the mean-field Hamiltonian $\hat{H}^{\text{mf}}$ is given by
\begin{equation}
\hat{H}^{\text{mf}} = \sum_{\alpha=x,y,z}J_{\alpha}\langle\hat{\sigma}^{\alpha}\rangle\hat{\sigma}^{\alpha},
\end{equation}
with $\langle\hat{\sigma}^{\alpha}\rangle = \text{Tr}(\hat{\sigma}^{\alpha}\hat{\rho})$.
By performing the integral on Eq.(\ref{MFmasterequation}), the self-consistent master equation will finally converge to a asymptotic steady state.
It is easy to find that the paramagnetic state with all the spins pointing down to the $z$-direction,  $\hat{\rho}_{\downarrow}=\bigotimes_{j}\hat{\rho}_{j,\downarrow}$, where $\hat{\rho}_{j,\downarrow}=|\downarrow_{j}\rangle\langle\downarrow_{j}|$ is always a steady-state solution to Eq.(\ref{MFmasterequation}).
\begin{figure}[!htp]
\includegraphics[width=0.50\textwidth]{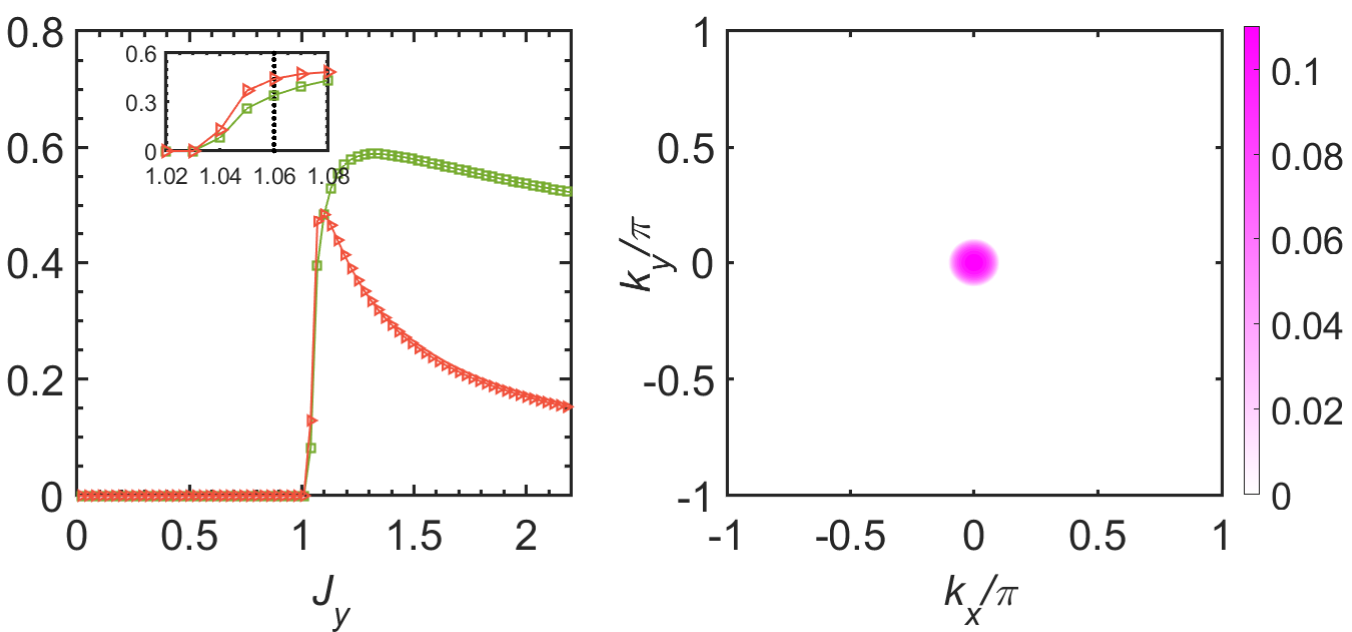}
\caption{\label{XYZmodelMF} Left panel: the mean-field steady-state magnetizations for $\langle\hat{\sigma}^{x}\rangle_{ss}$ (squares) and $\langle\hat{\sigma}^{y}\rangle_{ss}$ (triangles) as function of $J_{y}$. \xl{Right panel: the real part of the most unstable eigenvalue of the superoperator $\mathcal{L}_{\textbf{k}}$ in Eq. (\ref{Eq:LMatrix}) as a function of $k_{x}$ and $k_{y}$ for $J_y=1.06$. Other parameters are chosen as $J_{x} = 0.9$ and $J_{z} = 1$.}}
\end{figure}

Now we check the linear stability of the paramagnetic steady-state solution to the fluctuations \cite{MCCross1993RMP,AlexandreLeBoite2013PRL,AlexandreLeBoite2014PRA,XingLi2021PRB}. The fluctuations are added to each site as follows,
\begin{equation}
\hat{\rho} = \bigotimes_{j}(\hat{\rho}_{j} + \delta\rho_{j}).
\label{perturbed_DM}
\end{equation}
By performing the Fourier transform on the fluctuations $\delta \rho^{\textbf{k}}_{j} = \sum_{\textbf{k}}e^{-i\textbf{k}\cdot\textbf{r}_{j}}\delta \rho_{j}$ and substituting the perturbed density matrix in Eq. (\ref{perturbed_DM}), one can decouple the master equation (\ref{MFmasterequation}) in the momentum space as $\partial_{t}\delta\rho^{\textbf{k}}=\mathcal{L}_{\textbf{k}}\cdot\delta\rho^{\textbf{k}}$. The superoperator $\mathcal{L}_{\textbf{k}}$ reads,
\begin{equation}
\mathcal{L}_{\textbf{k}} =
\begin{pmatrix}
     -\gamma & 0                   & 0                    & 0 \\
      0      & P -\frac{\gamma}{2} & Q                    & 0 \\
      0      & -Q                  & -P -\frac{\gamma}{2} & 0 \\
      \gamma & 0                   & 0                    & 0
\end{pmatrix},
\label{Eq:LMatrix}
\end{equation}
where the coefficients are $P = -i[(J_{x} + J_{y})t_{\textbf{k}} - 2\mathfrak{z}J_{z}]$, $Q = -i(J_{x} - J_{y})t_{\textbf{k}}$, $\mathfrak{z}=4$ is the coordination number of two-dimensional square lattice, the vector is given by $t_{\textbf{k}} = 2\cos(k_{x}a) + 2\cos(k_{y}a)$ and $a$ is the lattice constant. The stability of the paramagnetic steady state can be revealed by the eigenvalue spectrum of superoperator $\mathcal{L}_{\textbf{k}}$. If the real parts of all the eigenvalues of $\mathcal{L}_{\textbf{k}}$ are negative, the system is stable under perturbation; otherwise, the system is unstable. Meanwhile, the critical points of PM-FM phase transition can be analytically determined by the eigenvalues of the Jacobian. the Jacobian is obtained by the system of nonlinear Bloch equations \cite{JiasenJin2016PRX,XingLi2021PRB}. The phase boundary can be expressed as follows,
\begin{equation}
J_{x, y}^{c}=\frac{1}{16\mathfrak{z}^2}\frac{1}{J_z-J_{y,x}}+J_z.
\label{CriticalPointXYZ}
\end{equation}

In the left panel of Fig. \ref{XYZmodelMF}, we show the steady-state magnetizations of in the $xy$ plane as a function of $J_{y}$. One can find that for $J_y<J^{(c)}_{y}\approx1.0391$, the steady-state magnetizations are $\langle\sigma^x\rangle_{ss}=\langle\sigma^y\rangle_{ss}=0$. While when $J_y$ go across the critical point $J^{(c)}_{y}$,  the magnetizaitions in the $xy$ plane become nonzero indicating the appearance of a continuous phase transition from the disordered PM phase to the ordered FM phase with $\mathbb{Z}_{2}$ symmetry breaking. In particular, for $J_y>J^{(c)}_{y}$, the state $\hat{\rho}_{\downarrow}$ is unstable to the uniform spatial perturbations as shown in the right panel of Fig. \ref{XYZmodelMF}. One can see that the system is mostly unstable to the perturbations with wave vector $\textbf{k} = (0, 0)$ and will be eventually driven to the FM phase. For the case of $J_{x} = 0.9$ and $J_{z} = 1$, under the MF approximation, the critical point can be analytically determined by Eq.(\ref{CriticalPointXYZ}).

\begin{figure}[!htp]
\includegraphics[width=0.49\textwidth]{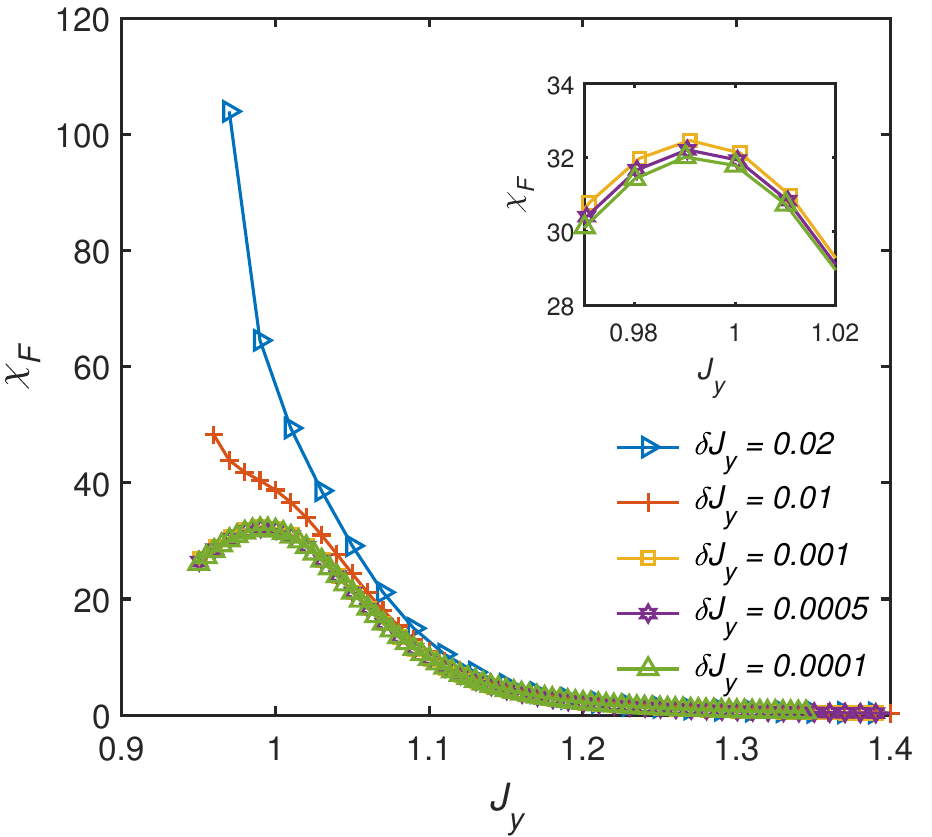}
\caption{\label{errordisplay} \xl{In different scales of perturbations $\delta J_{y}$, the fidelity susceptibility $\chi_F $ as a function of $J_{y}$ for a $2\times2$ square lattice. The parameters are chosen as $J_{x} = 0.9$ and $J_{z} = 1$. The inset shows the zoom-in for the curves corresponding to $\delta J_{y}\le 10^{-3}$.}}
\end{figure}

Now we investigate the susceptibilities of fidelity and trace distance in the vicinity of phase transition by varying the coupling strength. The amount of variation $\delta J_y$ should be sufficiently small by definition. In Fig. \ref{errordisplay} we have checked the convergence of fidelity susceptibility which computed with various magnitudes of $\delta J_y$. It is shown that the fidelity susceptibility starts to converge at $\delta J_{y} = 10^{-3}$. In the rest work of this model, unless otherwise stated, the parameter perturbation is fixed as $\delta J_{y} = 10^{-3}$.

\begin{figure}[!htp]
\includegraphics[width=0.49\textwidth]{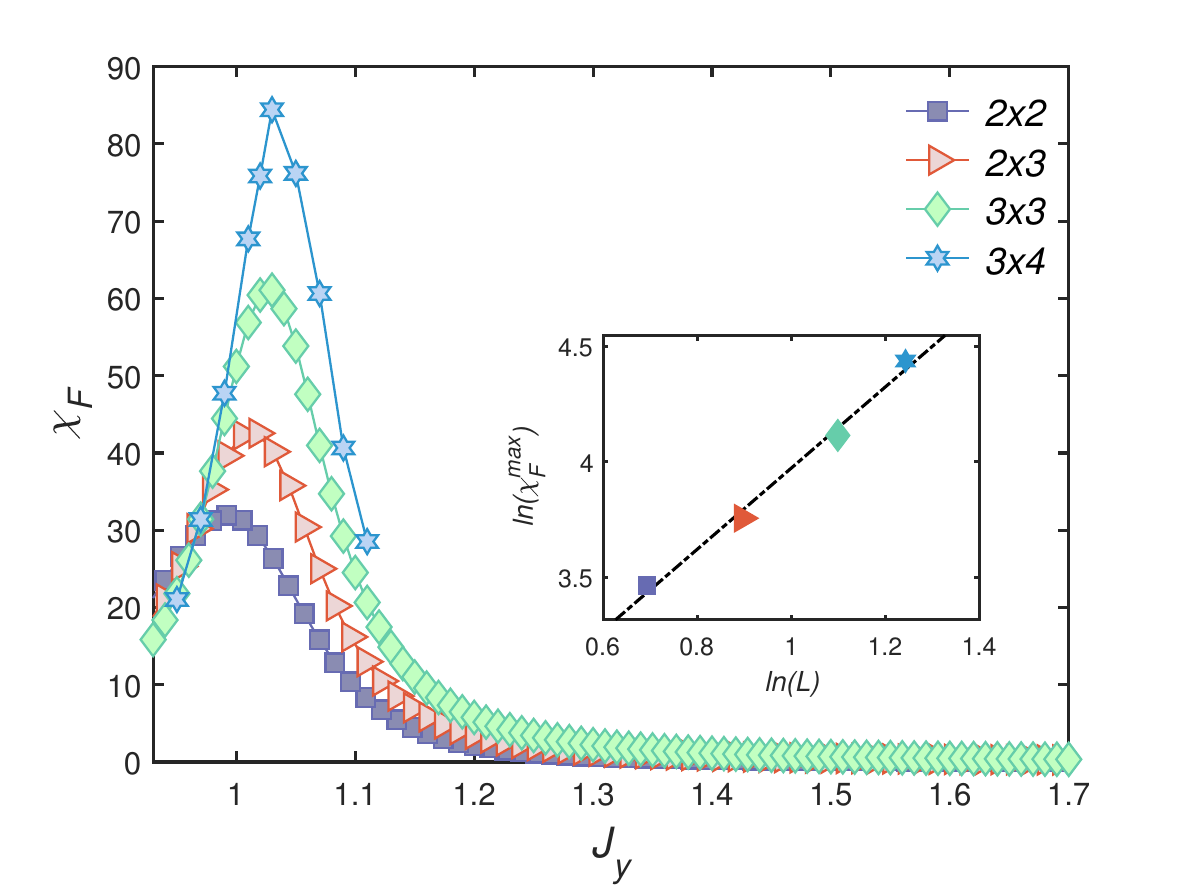}
\caption{\label{XYFFS} \xl{The fidelity susceptibility $\chi_F $ as a function of $J_{y}$ for $2\times2$, $2\times3$, $3\times3$ and $3\times4$ lattice, with the fixed $J_{x} = 0.9$ and $J_{z} = 1$. Inset: the maximum of fidelity susceptibility $\chi_F^{\it{max}}$ versus the linear dimension $L$ (log-log scale) with a power-law fitting \jj{(dash-dotted line)}.}}
\end{figure}

With the periodic boundary condition, the steady states are obtained in the followings manners: for the $2\times2$ and \xl{$2\times3$}  lattices, we exactly diagonalize the Liouvillian superoperator $\mathcal{L}$ and take the eigenstate associated to the zero eigenvalue as the steady state; while for $3\times3$ and $3\times4$ lattices, we numerically integrate the master equation via the fourth-order Runge-Kutta method and obtain the steady states by looking at the density matrix in the long-time limit, i.e. $\hat{\rho}_{ss} = \lim_{t\to+\infty} e^{\mathcal{L}t}\hat{\rho}(0)$.

\xl{
In Fig. \ref{XYFFS} we show the fidelity susceptibilities $\chi_F$ as functions of the coupling strength $J_y$ for the lattices in different sizes.
One can see that the maximum of $\chi_F$ always appears in the vicinity of $J_y=1$, indicating the abrupt change for the steady-state density matrices. Moreover, the peak of $\chi_F$ becomes sharper as the size of lattice increasing. In the inset of Fig. \ref{XYFFS}, the scaling of the maximum fidelity susceptibility versus the lattice size is shown. We find that power-law dependence of the maxima on the size of lattice as follows,
\begin{equation}
\chi_{F}^{max}\sim \kappa L^{\eta},
\label{Eq:ScalingChiF}
\end{equation}
where \jj{$L=N^{1/d}$ is the linear dimension of the system, $d=2$ is the real dimension and $N$ is the number of sites. Then we obtain the corresponding fitting parameters are $\eta \approx 1.7572$ and $\kappa\approx 2.2162$ .}

\begin{figure}[!htp]
\includegraphics[width=0.49\textwidth]{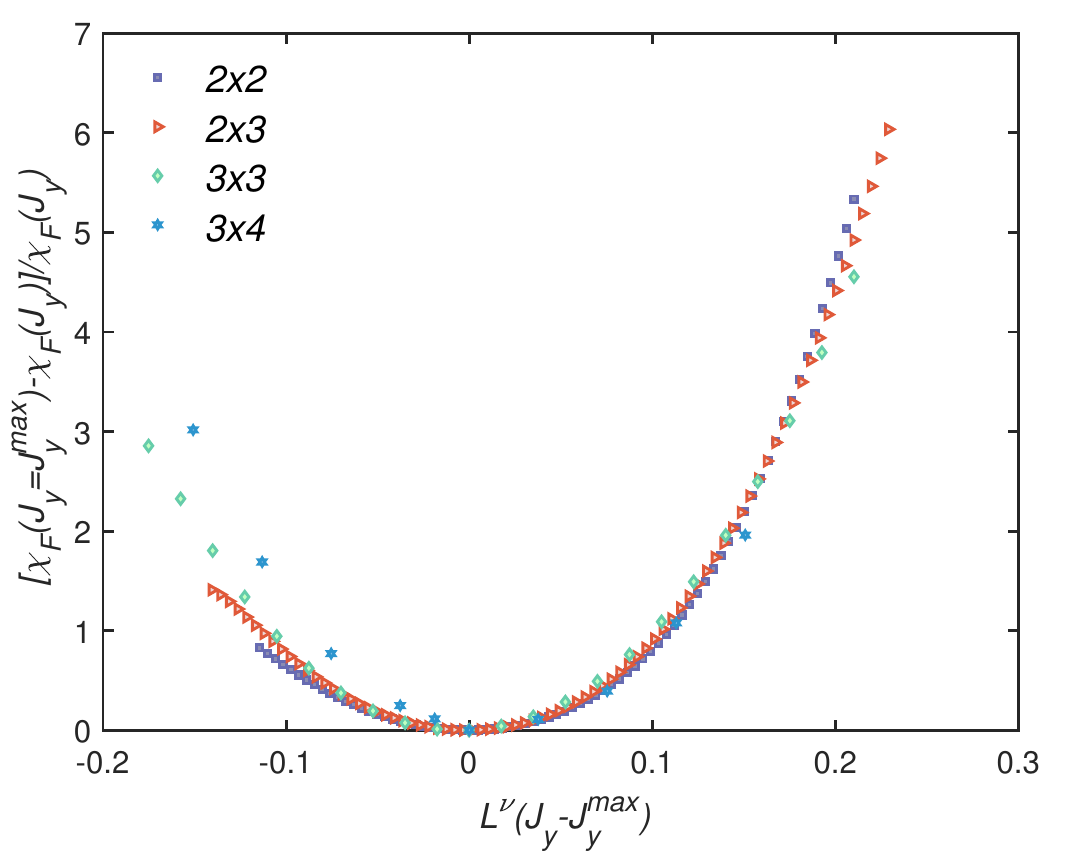}
\caption{\label{ScalingLaw}\xl{The rescaled fidelity susceptibility as a function of $L^{\nu}(J_{y}-J_{y}(\chi^{max}_{F}))$ in the finite-size scaling analysis of the two-dimensional dissipative XYZ model.
 The critical exponent for the correlation length is estimated to be $\nu=0.48\pm0.06$. The parameters are chosen as $J_x=0.9$ and $J_z=1$.}
 }
\end{figure}

Now we discuss our results within the scope of the scaling hypothesis. Suppose that, near the critical point $J^{c}_{y}$, the average fidelity susceptibility $\chi_{F}(J_{y},L)/L^d$ of a cluster of size $N=L^d$  scales as follows
\begin{equation}
\frac{\chi_{F}(J_{y},L)}{L^d}\sim \frac{1}{|J_{y}-J^{c}_{y}|^{\alpha}},
\end{equation}
where $\alpha$ is the corresponding exponent. Inspired by the proposals by Gu {\it et al.} \cite{GushijianPRB2008} and taking into account Eq. (18),
the rescaled fidelity susceptibility as a function of the rescaled coupling strength is given by
\begin{equation}
\frac{\chi_{F}(J_{y}(\chi^{max}_{F}),L)-\chi_{F}(J_{y},L)}{\chi_{F}(J_{y},L)}=f[L^{\nu}(J_{y}-J_{y}(\chi^{max}_{F}))],
\label{Eq:Scaling}
\end{equation}
where $\nu$ is the critical exponent of correlation length. In deriving Eq. (\ref{Eq:Scaling}), we have rewritten Eq. (18) as
$\chi_{F}(J_{y}(\chi^{max}_{F}),L)\sim L^{\eta}$ in which $J_{y}(\chi^{max}_{F})$ is the coupling strength for the maximal fidelity susceptibility in the lattices of different sizes. Consequently, we have the following relationship
\begin{equation}
\nu= \frac{\eta-d}{\alpha}.
\end{equation}
In Fig. \ref{ScalingLaw}, we show the rescaled fidelity susceptibilities with respect to the rescaled coupling strength. One can see that the data collapse with the estimated critical exponent \jj{$\nu=0.48\pm0.06$. Unfortunately, to the best of our knowledge, a direct estimation of critical exponent $\nu$ in the dissipative quantum XYZ model is not reported by any literature, we could not benchmark our estimation on $\nu$. Notice that the estimation is basing on the relative small-size lattices, the result should be interpreted with caution.}
}

\begin{figure}[!htp]
\includegraphics[width=0.49\textwidth]{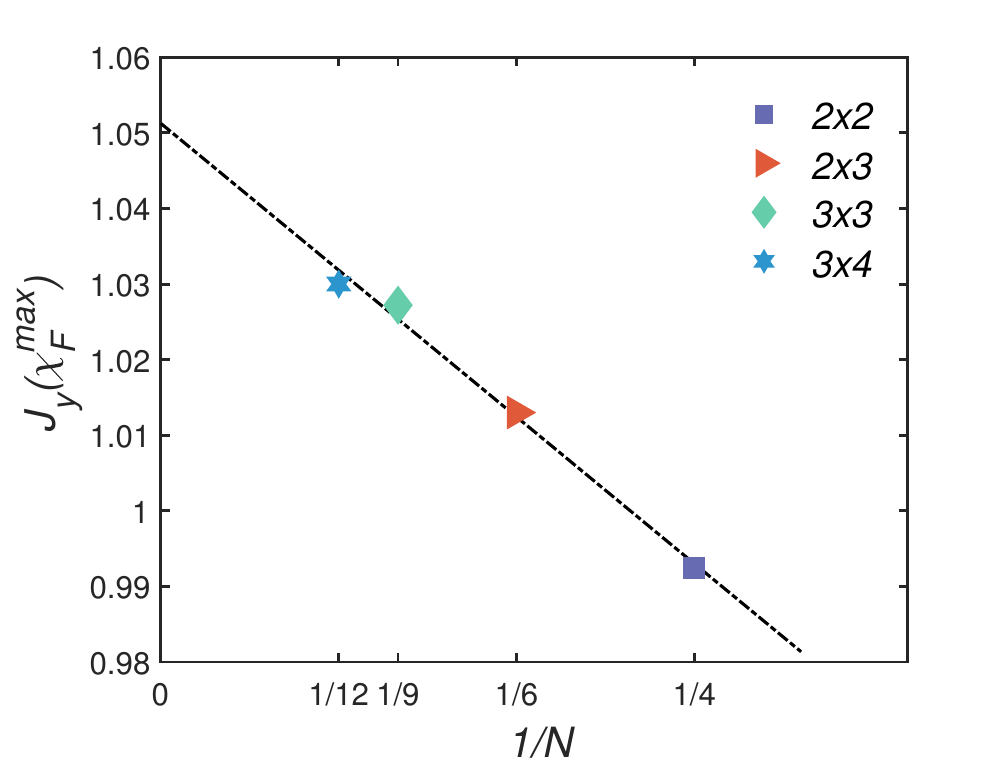}
\caption{\label{CriticalPoint} The coupling strengths corresponding to the \xl{maximum $\chi^{max}_{F}$} as a function of the different system sizes $1/N$. The \jj{dash-dotted} line is the linear fitting. The other parameters are $J_{x} = 0.9$ and $J_{z} = 1$.}
\end{figure}

In order to estimate the critical point $J^{c}_{y}$ of the phase transition, we linearly fit the the critical coupling strengths \xl{$J_{y}(\chi_{F}^{max})$}, which correspond to the extreme points of $\chi_{F}$, to the system size $N$. As shown in Fig. \ref{CriticalPoint}, with the system size increasing the critical coupling strength \xl{$J_{y}(\chi_{F}^{max})$} converge. In the the thermodynamic limit, i.e. $1/N\to0$, the critical coupling strength can be estimated as $J^{c}_{y}\approx1.05$. For comparisons, in the cluster mean-field and Gutzwiller Monte Carlo calculations the critical points are estimated to be $J^{c}_{y}\approx1.04$ \cite{JiasenJin2016PRX,Casteels2018}, by means of the quantum trajectory method the critical point is estimated to be $J^{c}_{y}\approx1.04$ \cite{Riccardo2018NJP}, and the corner-space renormalization predicts a phase transition at  $J^{c}_{y}\approx1.07$ \cite{Rota2017PRB}.

\begin{figure}[!htp]
\includegraphics[width=0.49\textwidth]{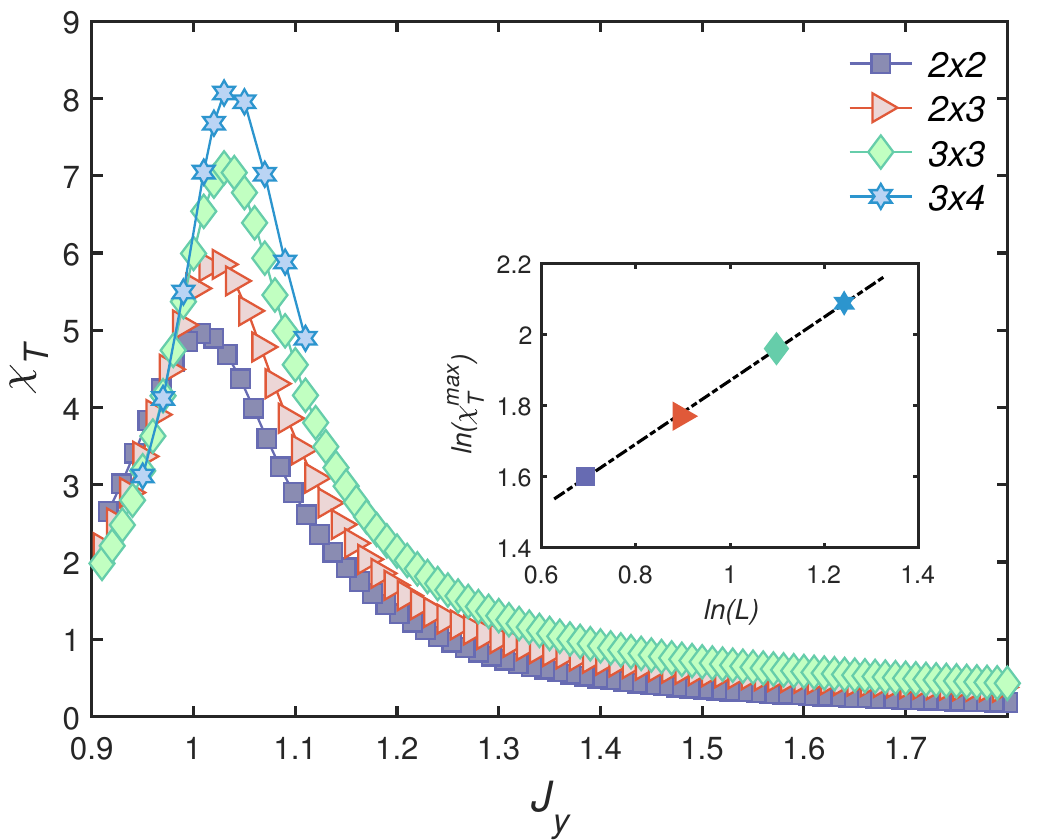}
\caption{\label{XYZTS} The trace distance susceptibility versus the coupling parameter $J_{y}$ for $2\times2$, $2\times3$, $3\times3$ and $3\times4$ lattices. \jj{The inset shows the maximum of the trace distance susceptibility $\chi^{max}_{T}$ versus the linear dimension $L$ (log-log scale) with a power-law fitting (dash-dotted line).} Other parameters are chosen as $J_{x} = 0.9$, $J_{z} = 1$ and $\delta J_{y} = 0.0001$.}
\end{figure}

In Fig. \ref{XYZTS}, we show the trace distance susceptibility as functions of $J_y$. One can see that, as the lattice size $N$ increasing the $\chi_{T}$ reported a critical behavior which is similar to the fidelity susceptibility. Near the critical point of the PM-FM phase transition, the maximal value of  $\chi_{T}$ keeps growing as the size increasing. Although the peak values are not growing as fast as the results of the fidelity susceptibility, a power-law fit for the peaks of the $\chi_{T}$ and the lattice number still can be found \xl{$\chi_{T}^{max}\propto L^{\zeta}$ with $\zeta\approx 0.8921$}. By linearly fitting the $J^{max}_{y}(N)$ to the system size $N$, we obtained critical point in thermodynamic limit as $J^{c}_{y}\approx 1.05$. This result is in good agreement with the existed results \cite{JiasenJin2016PRX,Casteels2018,Rota2017PRB,Riccardo2018NJP}.

\subsection{The driven-dissipative Kerr model}
\label{ModelII}

In this section, we concentrate on a driven-dissipative Kerr model, as an example of second-order dissipative phase transition with symmetry breaking in the continuous variable (CV) system. The specific model we investigate is a typical nonlinear oscillator with two-photon pumping but the single-photon dissipation. This model has already been studied in Ref.\cite{ZhangPRA2021} where the authors investigated the properties of continuous phase transition by means of mean-field theory, exact diagonalization, and the Keldysh formalism.

Here we briefly review the previous work studied by Zhang {\it et. al.}\cite{ZhangPRA2021}. In a rotating frame, the considered Hamiltonian can be obtained as,
\begin{equation}
\hat{H} = -\Delta\hat{a}^{\dagger}\hat{a} + \frac{U}{2}\hat{a}^{\dagger}\hat{a}^{\dagger}\hat{a}\hat{a} + \frac{G}{4}(\hat{a}^{\dagger}\hat{a}^{\dagger} + \hat{a}\hat{a}),
\label{Ham2}
\end{equation}
where $\Delta=\omega_{p}-\omega_{c}$ is the detuning of the frequency of pumping $\omega_{p}$ and cavity $\omega_{c}$, $U$ quantifies the Kerr nonlinearity and $G$ describes the amplitude of the two-photon driving. The specific single-photon dissipation is described by
\begin{equation}
\mathcal{D}[\hat{\rho}] = \frac{\gamma}{2}(2\hat{a}\hat{\rho}\hat{a}^{\dagger} - \{\hat{a}^{\dagger}\hat{a},\hat{\rho}\}).
\label{diss2}
\end{equation}
where $\gamma$ is the decay rate. Again we will work in units of $\gamma$. Substituting the Hamiltonian (\ref{Ham2}) and the dissipator (\ref{diss2}) into Eq. (\ref{MasterEquation}), one can find a discrete $\mathbb{Z}_{2}$ symmetry associated to the master equation. The corresponding symmetry superoperator is given by
\begin{equation}
\mathcal{Z}_{2}\bullet = e^{i\pi\hat{a}^{\dagger}\hat{a}}\bullet e^{-i\pi\hat{a}^{\dagger}\hat{a}},
\end{equation}
where the symbol $\bullet$ denotes the steady-state density matrix.

\begin{figure}[!htp]
\includegraphics[width=0.5\textwidth]{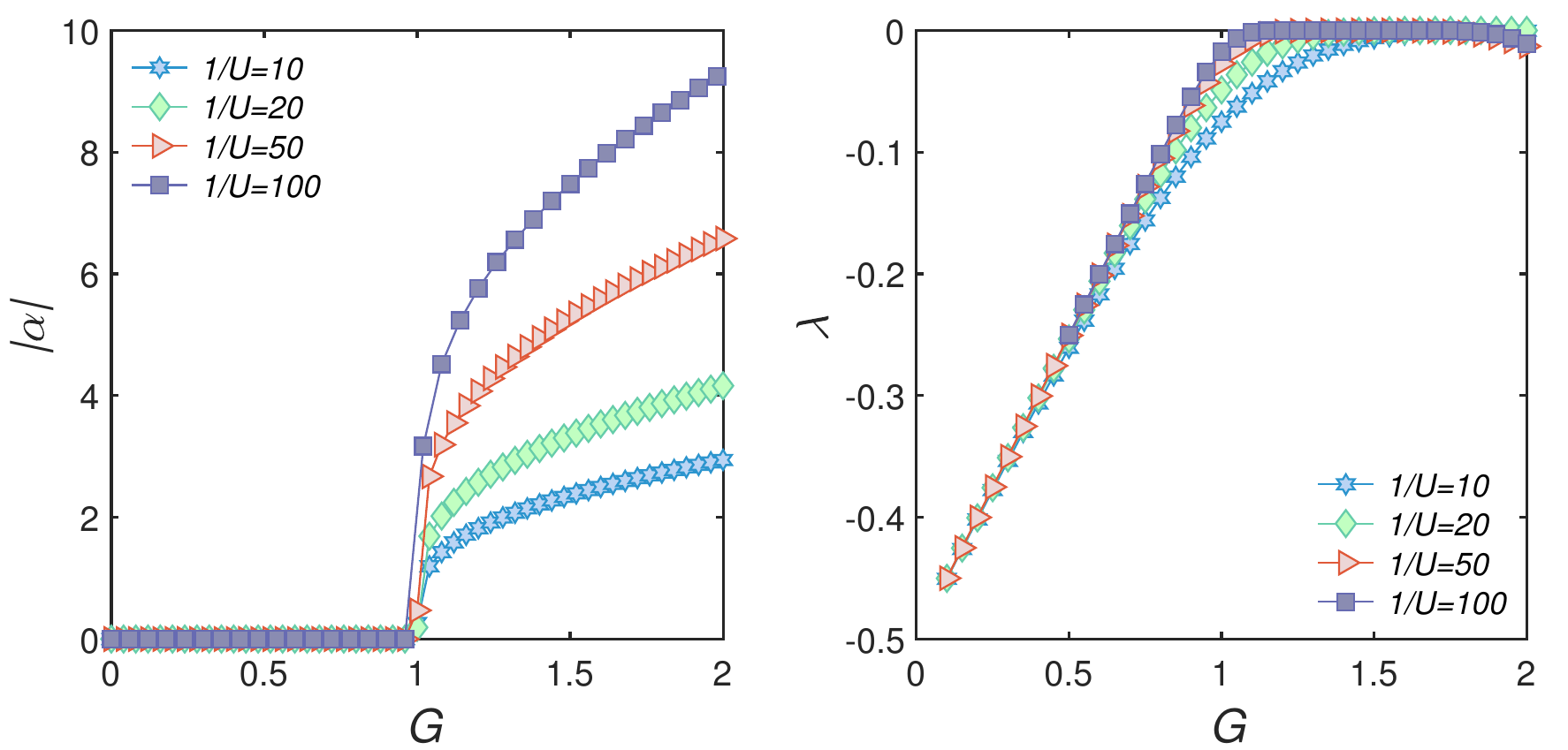}
\caption{\label{kerrMF} Left panel: The modulus of the steady-state coherent field amplitude $|\alpha|$ verse the two-photon driving $G$ under the semiclassical approximation treatment for the different Kerr nonlinearities. Right panel: The Liouvillian gap $\lambda = \text{Re}[\lambda_{1}]$ as a function of two-photon driving strength $G$ with the different Kerr nonlinearity.}
\end{figure}

Under the semiclassical approximation, in which all the quantum fluctuations and quantum correlations become negligible, i.e. $\langle \hat{a}^\dagger \hat{a}\hat{a} \rangle \approx \langle \hat{a}^\dagger\rangle\langle \hat{a}\rangle\langle \hat{a}\rangle$, we can obtain the equation of motion for the coherent field amplitude $\alpha=\langle \hat{a}\rangle$ in the resonant case ($\Delta=0$) \cite{CiutiRMP2013}
\begin{equation}
\frac{d}{dt} \alpha = (- iU |\alpha|^2 -\frac{\gamma}{2})\alpha - i\frac{G}{2}\alpha^*.
\label{EOM}
\end{equation}
The steady-state value of $\alpha$ can be obtained by numerically evolving the self-consistent equation of motion (\ref{EOM}) to a sufficiently long time. The steady-state value of $|\alpha_{\text{ss}}|$ is considered as the order parameter. Namely, the zero $|\alpha_{\text{ss}}|$ indicates the disordered phase while the nozero $|\alpha_{\text{ss}}|$ indicates the ordered phase with $\mathbb{Z}_2$ symmetry breaking. By analyzing the semiclassical equation (\ref{EOM}), the number of photons $n$ in the cavity is of order $\gamma/U$, and the thermodynamic limit can be achieved in the limit of infinitesimal interaction, $U/\gamma\to0^{+}$ \cite{ZhangPRA2021}.

In the left panel of Fig.\ref{kerrMF}, the modulus of the steady-state coherent field amplitude as a function of the two-photon driving strength $G$ is shown. For different values of the Kerr nonlinearity strength, the order parameter $|\alpha|$ always shows a transition from the zero value to a finite constant, implying the emergence of a second-order phase transition. At the level of the semiclassical approximation, the critical point of this dissipative quantum phase transition is near $G_{c}\approx1$.

Now we go beyond the semiclassical approximation and concentrates on the quantum level. We show the Liouvillian gap $\lambda = \text{Re}[\lambda_{1}]$ in the right panel of Fig.\ref{kerrMF}. One can see that with the Kerr nonlinearity decreasing, the Liouvillian gap tends to zero faster. For the minimum Kerr nonlinearity strength $U=1/100$, the Liouvillian gap closes in the interval of $G\in(1,2)$. This implies that the two degenerate steady states break the $\mathbb{Z}_2$ symmetry spontaneously characterizing the occurrence of the second-order phase transition \cite{Minganti2018}.

\begin{figure}[!htp]
\includegraphics[width=0.49\textwidth]{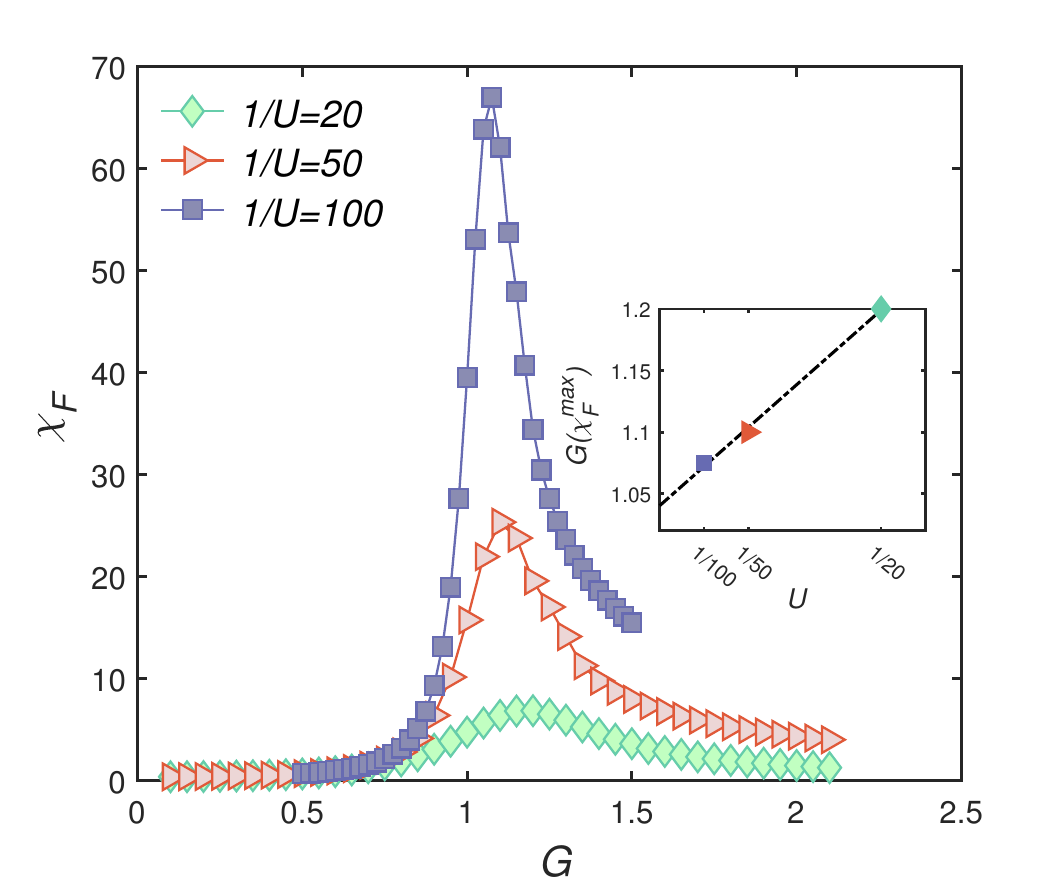}
\caption{\label{KerrFS} \xl{The fidelity susceptibility $\chi_{F}$ of the single Kerr oscillator versus the two-photon driving strength $G$. \jj{The different markers
indicate that the results are simulated in the different number of photons in the cavity. The inset shows the maximum value $\chi_F$ versus Kerr nonlinearity $U$, and the dash-dotted line indicates the finite-size linear fitting.}}}
\end{figure}

\xl{
In Fig.\ref{KerrFS}, we show the numerics of the fidelity susceptibility $\chi_{F}$ changes with the two-photon driving $G$. The different markers label the results of the corresponding Kerr nonlinearity strengths $U$. Analogous to the spin model discussed in Sec. \ref{ModelI}, the abrupt changes for the steady-state density matrices contribute to the peak patterns. As the number of photons increases, the maximum of $\chi^{max}_{F}$ keeps being higher and the peak is getting sharper. Moreover, with decreasing nonlinearity $U$ to the thermodynamic limit,  the critical two-photon driving $G(\chi_{F}^{max})$ shift towards the left. In the inset of Fig.\ref{KerrFS}, we report the scaling of $G(\chi_{F}^{max})$ with the Kerr nonlinearity $U$. Through the linear fitting one can approximately estimate the critical value as $G^{c}\approx 1.04$.}

\begin{figure}[!htp]
\includegraphics[width=0.49\textwidth]{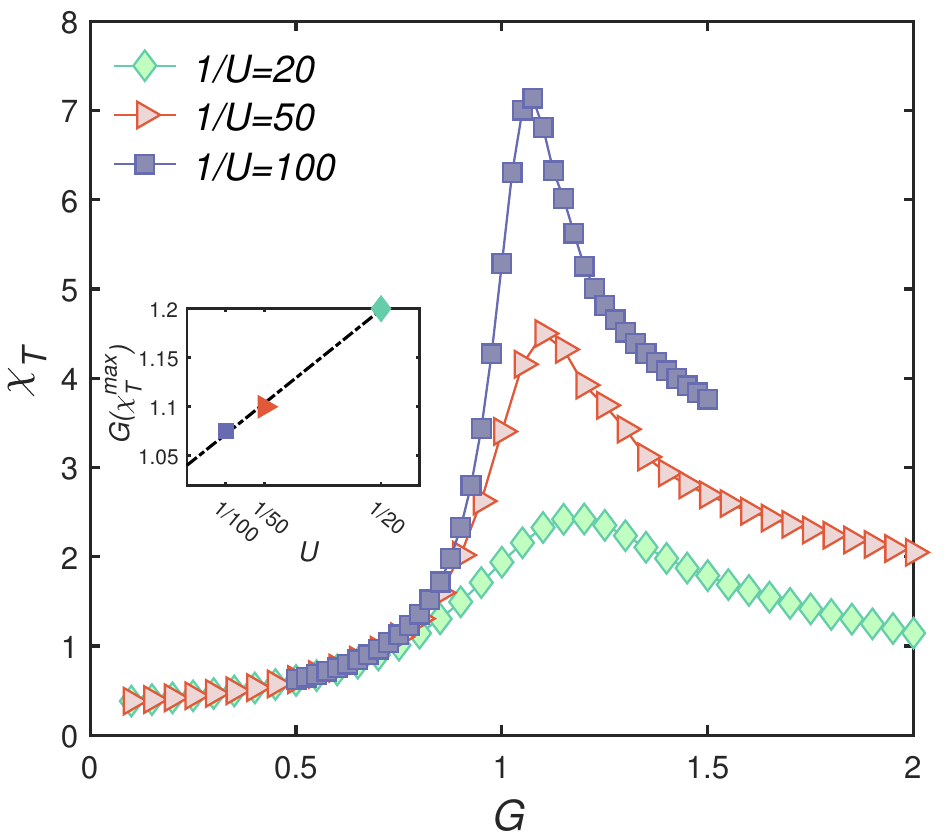}
\caption{\label{KerrTS} The trace distance susceptibility $\chi_{T}$ of the single Kerr oscillator versus the two-photon driving strength $G$. The different markers indicate that the results are simulated in the different number of photons in the cavity. \jj{The inset shows the linear fitting of maximum value $\chi_T$ to Kerr nonlinearity $U$, and the dash-dotted indicates the finite-size linear fitting.}}
\end{figure}

The divergence can also be observed in the behavior of trace distance susceptibility$\chi_{T}$ as shown in Fig. \ref{KerrTS}. Following the same analysis routine for $\chi_{F}$, we find that $\chi_{T}$ becomes divergent as the two-photon driving strength $G$ approaching to the critical point. As the Kerr nonlinearity $U$ decreasing, the divergent behavior of $\chi_T$ becomes more and more apparent, i.e. the height of the peaks become higher and the positions (critical values of two-photon driving $G_{\chi_T^{max}}$) shift towards left. The linear fitting of $G_{\chi_T^{max}}$ to the nonlinearity allow us to extrapolate the critical value of two-photon driving to be $G^{c}\approx1.04$, in consistent with the results obtained by the fidelity susceptibility.

\section{Summary}
\label{Summary}
In summary, we have utilized the susceptibilities of the fidelity and trace distance to detect the steady-state phase transitions in dissipative quantum systems. Different from the previous studies for dissipative phase transitions based on appropriately chosen order parameters, the two indicators proposed in this paper are observable-independent and, more interestingly, are the direct reflections of the abrupt changes of similarities between the steady states when the system undergoes a phase transition.

As applications, we mainly investigated the dissipative spin-1/2 XYZ model on two-dimensional square lattice and a driven-dissipative Kerr oscillator. It has been shown that both the two models may undergo continuous steady-state phase transitions (breaking $\mathbb{Z}_2$ symmetry) via tuning the controlling parameters.

In the former model, we first confirmed the existence of continuous phase by means of the mean-field approximation and the linear stability analysis. Then we studied behaviors of fidelity susceptibility and trace distance susceptibility of steady states as functions of coupling strength. We found the divergent behaviors of the susceptibilities near the phase transitions which stem from the abrupt change of the state similarity between two steady states when the systems undergo the phase transition. \jj{We performed the finite-size scaling analysis on the fidelity susceptibility. Within the scope of the scaling hypothesis, we estimated the values of the corresponding critical exponents. Moreover,} the finite-size scaling of the critical coupling strength and the system size enable us to estimate the true critical point in the thermodynamic limit. In the latter CV model, we revisited the steady-state properties obtained by the semiclassical treatment. Analogous to the spin model, the singular behaviors of susceptibilities as functions of the two-photon driving strength were observed which indicates the occurrence of dissipative phase transitions. In particular, the scaling of the critical driving strength to the Kerr nonlinearity allows us extrapolate the critical point in thermodynamic limit ($U\rightarrow0$). The critical points, in thermodynamic limit, accessed from the analysis on the fidelity and trace distance susceptibilities agree well with the existed results obtained by other methods.

Finally, the investigation of the dissipative quantum phase transition is still a hard task, especially for the quantum many-body systems. The exponential growth of the quantum many-body Hilbert space restricts the capacities of directly witness the phenomena of phase transitions. Fortunately, the critical behaviors of the fidelity susceptibility, trace distance susceptibility, angularly averaged magnetic susceptibility and quantum fisher information can indirectly reveal the existence of the phase transitions. Along this line, the combination of the fidelity susceptibility and trace distance susceptibility with other state-of-the-art simulation strategies for exploring larger lattice is promising, such as the matrix-product-operator approach \cite{MascarenhasPRA2015}, the neural networks \cite{MichaelJHartmann2019PRL,AlexandraNagy2019PRL,FilippoVicentini2019PRL,NobuyukiYoshioka2019PRL,DengarXiv2020}. Future exploration of other physical models and phenomena, e.g., geometrical frustration \cite{XingLi2021PRB,lxlarXiv2021,JingQian2013PRA,ZejianPRA2021} is also an intriguing perspective.

\section*{ACKNOWLEDGMENTS}
This work is supported by the National Natural Science Foundation of China via Grant No. 11975064.

\end{document}